\documentstyle[12pt]{article}
\begin{document}
\author{G.Palma and J.C. Rojas \\
Departamento de F\'\i sica,Universidad de Santiago de Chile,\\
Casilla 307, Correo 2, Santiago de Chile.}
\title{Variational Study of the Phase Transition at Finite T in the $\lambda
\phi^4 $-Theory }
\date{June 8, 1995 }
\maketitle

\vspace{2.0cm}

\newpage

\section{Introduction}

In recent years the question of the nature of the phase transitions in
Quantum Field Theories has been a long standing subject of study, specially
in gauge theories at finite temperature, because of its cosmological
implications on the Baryonic asymmetry of the universe \cite{linde}.

In spite of a large amount of work done in this subject \cite{wetterich}, we
do not have a conclusive answer about the nature of the phase transition
even in toy models like the scalar electrodynamics at finite temperature.
The numerical approach has been extremely elusive, and we are still waiting
for trustable predictions which are not sensitive to the method used. At $%
T=0 $ the conventional loop expansion fails to predict spontaneous symmetry
breaking in the scalar $\lambda \Phi ^4$ theory: the non-trivial minimum
lies outside the validity region of the one-loop expansion \cite{dolan}. In
addition the re-summation of certain classes of diagrams, daisy or
super-daisy diagrams does not improve the situation: a first order phase
transition is found in contradiction with the evidence supported by a large
amount of numerical as well as analytical renormalization group studies\cite
{luescher}.

We propose an alternative method based on the triviality of the 4 -
dimensional $\lambda \Phi ^4$theory and the use of the so called
Feynman-Bogoliubov method \cite{tyablikov}, which consists in the solution
of a gap equation to compute the effective potential non-perturbatively \cite
{ute}. To implement this, we consider the $\lambda \Phi ^4$ theory as a
function of a finite cut off $\Lambda $ lying within the scaling region, or
area in the phase diagram where the low energy physical amplitudes depends
only weakly on the cutoff, and apply a variational method to compute the
effective potential.

One obtains an expression for the effective potential which depends on the
plasma mass, which is given self-consistently by a gap-equation. This
approach should be specially accurate very close to the phase transition as
it is shown in field theory at zero dimensions \cite{victor}. Its
perturbative solutions come from summations of certain classes of superdaisy
diagrams \cite{buchmuller}.

We analyze numerically the shape of the effective potential to obtain
information of the phase transition. The cutoff is determined by an
stability condition on the gap equation for the plasma mass and it is
expressed by a simple function of the bare parameters.

\section{\bf The Method}

We start with the Euclidean action of the one-component $\lambda \phi ^4$
theory expressed as a function of the fluctuation field $\zeta $, which
represents the high frequency component of the fundamental Higgs field $\phi
$ and has no zero-momentum Fourier component:
$$
S\left[ \zeta \right] =\int_{}^{}dz\left[ \frac 12\left( \partial _\mu \zeta
\left( z\right) \right) ^2+\frac 12m_{}^2(\Phi )\zeta ^2\left( z\right)
\right.
$$

\begin{equation}
\label{action}\left. \quad \quad \quad \quad \quad +\frac \lambda {4!}\left(
4\Phi \zeta ^3\left( z\right) +\zeta ^4\left( z\right) \right) +U_{cl}\left(
\Phi \right) \right]
\end{equation}

\noindent where $U_{cl}$ is the classical potential

$$
U_{cl}(\Phi )=\frac 12m_o^2\Phi ^2+\frac \lambda {4!}\Phi ^4\quad \quad
,\quad \quad m_{}^2(\Phi )=U_{cl}^{"}(\Phi )
$$

The constraint effective potential $U_{eff}(\Phi )$ for finite volume V
gives the probability distribution of the magnetization $\Phi =\frac 1V\int
\phi \left( z\right) dz$ and is defined as the delta constrained path
integral \cite{fukuda}:

\begin{equation}
\label{defuef}
\begin{array}{ll}
e^{-VU_{eff}(\Phi )} & =\int_{}^{}D\zeta \ \delta (C\zeta )e^{-S\left[ \zeta
\right] }
\begin{array}{cc}
&  \\
&  \\
&
\end{array}
\\
\begin{array}{cc}
&  \\
&  \\
&
\end{array}
& =e^{-VUcl(\Phi )}\int_{}^{}d\mu _\Gamma (\zeta )\ \exp \left[ -\int dz%
\frac{%
\begin{array}{c}
\\
\end{array}
\lambda }{%
\begin{array}{c}
\\
\end{array}
4!}\left( 4\Phi \zeta ^3+\zeta ^4\right) \right]
\end{array}
\end{equation}

\noindent where $d\mu _\Gamma (\zeta )$ is the gaussian measure with
covariance $\Gamma $ , the propagator of the fluctuation field $\zeta $. If $%
C$ denotes the averaging operator an $C^{\dagger }$ its adjoint, we write
explicitly:

$$
d\mu _\Gamma (\zeta )=\exp -\frac 12(\zeta ,\Gamma ^{-1}\zeta )\ D\zeta
\hspace{1cm} ;\hspace{1cm} \Gamma =\lim_{\kappa \to \infty }%
(-\bigtriangleup +\kappa C^{\dagger }C)^{-1}
$$

\noindent The limit $\kappa \rightarrow \infty $ can be taken if desired or
kept $\kappa $ finite for lattice computations \cite{hasenfratz}.

The expression (\ref{defuef}) has the adequate form to apply the
Feynman-Bogoliubov method to find the best quadratic approximation $%
S_o\left[ \zeta \right] =\frac 12(\zeta ,J\zeta )$ around which to expand.
The optimal quadratic approximation $J$ is determined by the extreme
condition on the Peierls inequality \cite{ute2}:

$$
\left\langle \frac{\delta ^2S\left[ \zeta \right] }{\delta \zeta (z_1)\delta
\zeta (z_2)}\right\rangle _0=J(z_1,z_2)
$$

\noindent Herein, $<..>_o$ is the expectation value in the theory with
optimal action $S_o\left[ \zeta \right] $. This is equivalent to the self
consistent equation :

\begin{equation}
\label{gapeq}J(z_1,z_2)=\Gamma ^{-1}(z_1,z_2)+\delta (z_1-z_2)\left\{
m^2(\Phi )+\frac \lambda 2J^{-1}(z_1,z_1)\right\}
\end{equation}

\noindent We seek for a translational invariant solution: $%
J(z_1,z_2)=J(z_1-z_2)$. Inserting eqn.(\ref{gapeq}) into the r.h.s. Peierls
inequality, we obtain the constraint effective potential:

\begin{equation}
\label{uef2}
\begin{array}{ll}
U_{eff}^{FB}(\Phi ) & =U_{cl}(\Phi )+\frac 1VTr\ln J+\frac 12m^2(\Phi
)J^{-1}(0)
\begin{array}{cc}
&  \\
&  \\
&
\end{array}
\\
\begin{array}{cc}
&  \\
&  \\
&
\end{array}
& +\frac 12\int dz_{}\ \ \Gamma ^{-1}(z)J^{-1}(z)+\frac \lambda 8\left(
J^{-1}(0)\right) ^2
\end{array}
\end{equation}

\noindent In the infinite volume limit this expression coincides with the
gaussian effective potential of ref \cite{stevenson}. His analysis amount to
discuss the triviality of the $\lambda \phi ^4$ - theory and find suitable
renormalization conditions in continuum in different dimensions.

The quadratic operator $J^{-1}(z_1,z_2)$ is given by:

$$
J^{-1}(z_1,z_2)=\int \frac{d^4p}{(2\pi )^4}\frac{e^{-ip(z_1-z_2)}}{%
p^2+m_p^2(\Phi )}
\begin{array}{c}
\\
\end{array}
$$
\noindent Thus the value of $m_p^2(\Phi )$ is given by the self consistent
gap equation:

\begin{equation}
\label{gap2}m_p^2(\Phi )=m_o^2(\Phi )+\frac \lambda 2\int \frac{d^4p}{(2\pi
)^4}\frac 1{p^2+m_p^2(\Phi )}
\begin{array}{c}
\\
\end{array}
\end{equation}

\noindent This gap equation has perturbative solutions which comes from
summation of superdaisy diagrams.

The introduction of a finite cutoff allows to compute explicitly the
integrals appearing in eqns.(\ref{uef2}) and(\ref{gap2}). We obtain the
final expression for $U_{eff}^{FB}(\Phi )$:

$$
U_{eff}^{FB}(\Phi )=U_{cl}(\Phi )+\ \frac 1{
\begin{array}{c}
\end{array}
64\pi ^2}\left\{ \Lambda ^4\ln \left( m_p^2(\Phi )+\Lambda ^2\right) -\frac{%
\Lambda ^4}2\right.
$$

$$
\left. \ \ \ \ \ \ \ \ \ \ \ \ \ \ \ \ \ \ \ \ \ +m_p^2(\Phi )\Lambda
^2-m_p^4(\Phi )\ln \left( 1+\frac{%
\begin{array}{c}
\end{array}
\Lambda ^2}{%
\begin{array}{c}
\end{array}
m_p^2(\Phi )}\right) \right\}
$$

\begin{equation}
\label{uef3}\ \ \ \ \ \ \ \ \ \ \ \ \ \ \ \ \ \ \ \ \ \ \ \ \ \ \ \ +\frac{%
\lambda
\begin{array}{c}
\end{array}
}{%
\begin{array}{c}
\end{array}
2048\pi ^2}\left\{ \Lambda ^4-2\Lambda ^2m_p^2(\Phi )\ln \left( 1+\frac{%
\begin{array}{c}
\end{array}
\Lambda ^2}{%
\begin{array}{c}
\end{array}
m_p^2(\Phi )}\right) \right.
\end{equation}

$$
\left. \ \ \ \ \ +m_p^4(\Phi )\ln \left( 1+\frac{%
\begin{array}{c}
\end{array}
\Lambda ^2}{m_p^2(\Phi )}\right) \right\}
$$

\noindent Here the plasma mass given by

\begin{equation}
\label{gap3}m_p^2(\Phi )=m_o^2(\Phi )+\frac \lambda {32\pi ^2}\left\{
\Lambda ^2-m_p^2(\Phi )\ln \left( 1+\frac{\Lambda ^2}{m_p^2(\Phi )}\right)
\right\}
\end{equation}

\section{Renormalization}

In order to compute the curves of constant physics or renormalization group
trajectories one has to find suitable physical quantities to keep constant
\cite{montvay}. For instance we use the renormalized coupling constants
which parametrize the action. We define the renormalized mass $m_R$ and self
coupling $\lambda _R$constant through the second and fourth derivative
respectively at the non trivial minimum in the broken phase to avoid
infrared divergencies, and at the origin in the symmetric phase.

The minimum of $U_{eff}^{FB}(\Phi )$ is given by the condition

\begin{equation}
\label{1deriv}\frac{dU_{eff}^{FB}(\Phi )}{d\Phi }=\Phi \left( m_p^2(\Phi
_{\min })-\frac \lambda 3\Phi _{\min }^2\right) =0
\begin{array}{c}
\\
\end{array}
\end{equation}

\noindent The explicit expression for $m_R^2$ and $\lambda _R$ can be given
in a closed form

\begin{equation}
\label{mrenorm}
\begin{array}{c}
\\
\\
\end{array}
m_R^2=\overline{m}_p^2-\lambda \Phi _{\min }^2+\frac{\lambda \Phi _{\min }^2%
}{1+\frac{\lambda
\begin{array}{c}
\end{array}
}{2
\begin{array}{c}
\end{array}
}I_2(\overline{m}_p^2)}
\end{equation}

$$
\lambda _R=-2\lambda +\frac{3\lambda
\begin{array}{c}
\end{array}
}{%
\begin{array}{c}
\end{array}
\left( 1+\lambda I_2(\overline{m}_p^2)/2\right) }
\begin{array}{c}
\\
\\
\end{array}
+\frac{%
\begin{array}{c}
\end{array}
6\lambda ^3\Phi _{\min }^2I_3(\overline{m}_p^2)}{%
\begin{array}{c}
\end{array}
\left( 1+\lambda I_2(\overline{m}_p^2)/2\right) ^3}.
\begin{array}{c}
\\
\\
\end{array}
$$

\begin{equation}
\label{lambdar}\ \ \ \ \ \ \ \ \ \ \ \ \ \ \ \ \ \ \ \ \ \ -\frac{%
\begin{array}{c}
\end{array}
3\lambda ^4\Phi _{\min }^4I_4(\overline{m}_p^2)}{%
\begin{array}{c}
\end{array}
\left( 1+\lambda I_2(\overline{m}_p^2)/2\right) ^4}
\begin{array}{c}
\\
\\
\end{array}
+\frac{%
\begin{array}{c}
\end{array}
3\lambda ^5\Phi _{\min }^4I_3^2(\overline{m}_p^2)}{%
\begin{array}{c}
\end{array}
\left( 1+\lambda I_2(\overline{m}_p^2)/2\right) ^5}
\begin{array}{c}
\\
\\
\end{array}
\end{equation}

\noindent where $\overline{m}_p^2=m_p^2(\Phi _{\min }^{})$ and $%
I_n(m_p^2)=\int \frac{d^4p}{(2\pi )^4}(p^2+m_p^2)^{-n}$ . In particular:

$$
I_2(m_p^2)=\frac 1{8\pi ^2}\left\{ \frac{-m_p^2}{2\left( m_p^2+\Lambda
^2\right) }-\frac 12\ln \left( 1+\frac{\Lambda ^2}{m_p^2}\right) \right\}
$$

\noindent and

$$
I_3(m_p^2)=\frac 1{32\pi ^2m_p^2}\quad ,\quad I_4(m_p^2)=\frac 1{96\pi
^2m_p^4}
$$

\section{Results and Discussion}

The starting point is to allocate in the space of the bare parameters, which
of course depend on the cutoff, the critical line where the renormalized
mass vanishes. This is known to be a non-perturbative problem. The
Feynman-Bogoliubov method is expected to describe very accurately the theory
very close to the transition line. the enlighten point to observe is that
the renormalized mass equals the plasma mass given by the gap equation when
the transition line is approached from the symmetric phase. Therefore we
seek the values of the bare parameters where the stability condition for the
gap equation is reached:

\begin{equation}
\label{extr0}\lambda _o=-32\pi ^2\frac{m_o^2}{\Lambda _{}^2}
\end{equation}

\noindent We use the lattice parametrization $m_{latt}^2=m_o^2a^2$ with $%
a=\pi /\Lambda $, and plot in figure 1 the equation (\ref{extr0}). A
remarkable correspondence is found when comparing with the more accurate
determination of the critical line by L\"uscher et al. \cite{luescher},
which is a meaningful test of the self - consistency of our approach that
transcend perturbation theory.

For an arbitrary cutoff which only ensures the validity of the stability
condition, figure 2 shows the shape of the effective potential computed from
equation (\ref{uef3}) for values of the bare parameter very close to the
critical value. A second order phase transition is clearly found as the non
trivial minimum evolves continuously to the origin of the desired field $%
\Phi $, as the critical line is approached. This result is independent of
the choice of the region in the bare parameter space provided they are close
enough to the critical line. This is in agreement with the large amount of
numerical evidence on the transition of the $\lambda \phi ^4$-theory at zero
temperature \cite{luescher}.

In figure 3 the renormalized coupling constant $\lambda _R$ is plotted as a
function of the cutoff $\Lambda $ for fixed values of the renormalized mass $%
m_R^{}$ and the lattice bare mass $m_{latt}^{}$. We conclude that it is not
possible to increase indefinitely the UV cutoff without entering an
instability region. This clearly suggests the existence of a triviality
bound for $\Lambda $ or equivalently the triviality of the lattice $\lambda
\phi ^4$-theory.

By choosing periodic boundary conditions in time direction with period $%
\beta =1/T$ , which leads to the well known sums on Matsubara frequencies $%
\omega _n=2\pi n/\beta $ one includes finite temperature effects \cite{ute2}%
. In this case we obtain a modified equation for the critical line at finite
temperature:

\begin{equation}
\label{extrt}\lambda _o=-\frac{m_{latt}^2}{\left( 1/32\pi ^2+T_B^2/24\Lambda
_{}^2\right) }
\end{equation}

\noindent Moreover from figure 4 we see that the symmetry restoration at
high temperature is driven by a weak first order phase transition. In this
situation the strength of the transition is sensitive to the value of the
cutoff.

Our concluding remarks about this method is to emphasize its simplicity and
beauty, and the remarkable agreement with numerical simulations much more
computer time consuming. Stimulated by this work, we are extending the
method to gauge theories.

\noindent {\bf Acknowledgments}

This work was supported in part by DICYT {\#} 049331PA and FONDECYT {\#}
1930067.

\end{document}